\documentclass[sigconf]{acmart}

\usepackage{soul}
\usepackage{xcolor}
\usepackage{xspace}
\usepackage{url}
\usepackage{graphicx}
\usepackage{listings}
\usepackage[caption=false]{subfig}  
\usepackage{amsmath}

\usepackage{amssymb}
\usepackage{textcomp}
\usepackage{breakurl} 
\usepackage{hyperref}
\usepackage{cleveref}
\usepackage{algorithm}
\usepackage{enumitem}
\usepackage{multirow}
\usepackage{color}
\usepackage{colortbl}
\usepackage{balance}

\newcommand{\gptO}{gpt-4o}

\newcommand{\gptOMini}{gpt-4o-mini}

\graphicspath{ {./img/} }

\copyrightyear{2026}
\acmYear{2026}
\setcopyright{cc}
\setcctype{by}
\acmConference[ICSE-SEET '26]{2026 IEEE/ACM 48th International Conference on Software Engineering}{April 12--18, 2026}{Rio de Janeiro, Brazil}
\acmBooktitle{2026 IEEE/ACM 48th International Conference on Software Engineering (ICSE-SEET '26), April 12--18, 2026, Rio de Janeiro, Brazil}
\acmPrice{}
\acmDOI{10.1145/3786580.3786976}
\acmISBN{979-8-4007-2423-7/2026/04}

\begin{document}

\title[Enhancing Debugging Skills with AI-Powered Assistance: A Real-Time Tool for Debugging Support]{Enhancing Debugging Skills with AI-Powered Assistance:\\A Real-Time Tool for Debugging Support}

\author{Elizaveta Artser}
\orcid{0009-0007-4919-2698}
\affiliation{
  \institution{JetBrains Research}
  \city{Munich}
  \country{Germany}
}
\email{elizaveta.artser@jetbrains.com}

\author{Daniil Karol}
\orcid{0009-0005-1189-901X}
\affiliation{%
  \institution{JetBrains Research}
  \city{Berlin}
  \country{Germany}}
\email{daniil.karol@jetbrains.com}

\author{Anna Potriasaeva}
\orcid{0009-0004-4686-2412}
\affiliation{
  \institution{JetBrains Research}
  \city{Belgrade}
  \country{Serbia}
}
\email{anna.potriasaeva@jetbrains.com}

\author{Aleksei Rostovskii} 
\orcid{0009-0008-8608-5186}
\affiliation{
  \institution{JetBrains Research}
  \city{Berlin}
  \country{Germany}
}
\email{aleksei.rostovskii@jetbrains.com}

\author{Katsiaryna Dzialets}
\orcid{0009-0003-3269-3622}
\affiliation{
  \institution{JetBrains Research}
  \city{Munich}
  \country{Germany}
}
\email{katsiaryna.dzialets@jetbrains.com}

\author{Ekaterina Koshchenko} 
\orcid{0000-0003-3375-037X}
\affiliation{
  \institution{JetBrains Research}
  \city{Amsterdam}
  \country{Netherlands}
}
\email{ekaterina.koshchenko@jetbrains.com}

\author{Xiaotian Su}
\orcid{0009-0004-0548-1576}
\affiliation{%
  \institution{ETH Zurich}
  \city{Zurich}
  \country{Switzerland}
}
\email{xiaotian.su@inf.ethz.ch}

\author{April Yi Wang}
\orcid{0000-0001-8724-4662}
\affiliation{%
  \institution{ETH Zurich}
  \city{Zurich}
  \country{Switzerland}
}
\email{april.wang@inf.ethz.ch}

\author{Anastasiia Birillo}
\orcid{0000-0003-2269-8211}
\affiliation{
  \institution{JetBrains Research}
  \city{Belgrade}
  \country{Serbia}
}
\email{anastasia.birillo@jetbrains.com}

\renewcommand{\shortauthors}{Elizaveta Artser et al.}

\begin{abstract}

Debugging is a crucial skill in programming education and software development, yet it is often overlooked in CS curricula. To address this, we introduce an AI-powered debugging assistant integrated into an IDE. It offers real-time support by analyzing code, suggesting breakpoints, and providing contextual hints. Using RAG with LLMs, program slicing, and custom heuristics, it enhances efficiency by minimizing LLM calls and improving accuracy. A three-level evaluation -- technical analysis, UX study, and classroom tests -- highlights its potential for teaching debugging.

\end{abstract}



\maketitle

\begin{figure*}[t]%
    \centering
    \subfloat[\centering Task panel with (1) Check button, \newline (2) failed test, (3) session-start message.]{\includegraphics[width=0.265\linewidth]{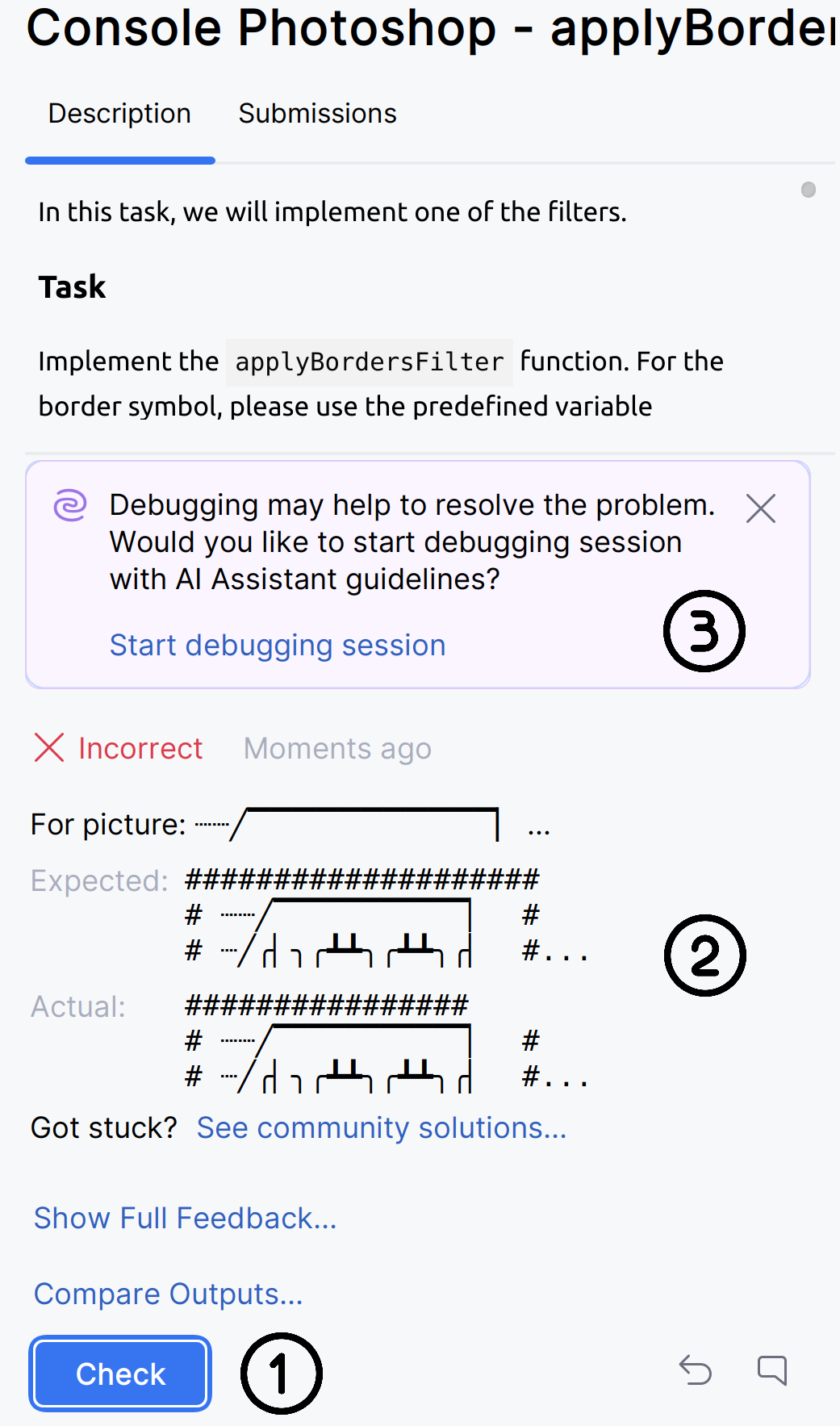}\label{fig:tool:start_debug}}
    \qquad
    \subfloat[\centering Code editor during a guided debugging session with (1) breakpoints suggested by the AI Debugging Assistant, (2)  breakpoints set by the student, and (3) breakpoint hint.]{\includegraphics[width=0.685\linewidth]{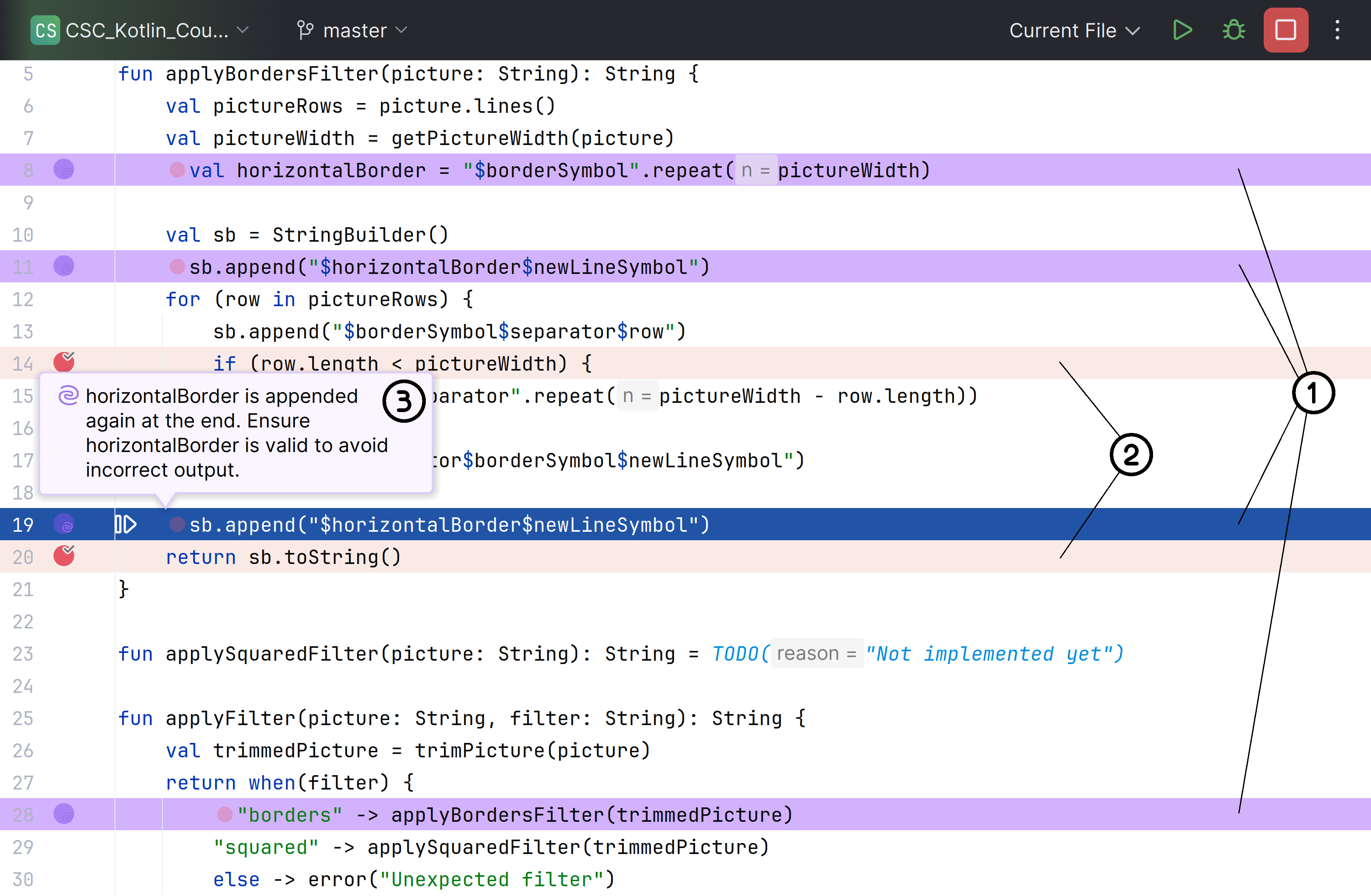}\label{fig:tool:breakpoints}}
    \caption{Main components of the AI Debugging Assistant tool.}
    \label{fig:tool:general}
\end{figure*}

\section{Introduction}
\label{section:introduction}

Debugging is an essential skill in programming education ~\cite{fitzgerald2009debugging} and professional software development~\cite{mullerburg1983role}, yet it is rarely taught explicitly in CS curricula~\cite{Ahmadzadeh2005}. 
Most studies focus on pre-designed exercises that teach debugging concepts~\cite{padurean2025bugspotter, wilkin2025debugging, garcia2022read, luxton2018ladebug, miljanovic2017robobug} but fail to help students apply these skills to real coding tasks~\cite{yang2025decoding}.
At the same time, providing real-time feedback can significantly improve students' performance and enthusiasm for learning~\cite{ma2024hypocompass, wiggins2021exploring, marwan2020adaptive}.

Recent advancements in Large Language Models (LLMs) offer potential solutions for providing real-time debugging assistance during programming exercises. LLMs are already widely used for program repair~\cite{kazemitabaar2024codeaid} and generating personalized feedback in education~\cite{ma2024hypocompass}. However, LLMs can generate inaccurate or unreliable responses, making it difficult to fully depend on them~\cite{smith2025spotting, koutcheme2025evaluating, chang2024survey, zhang2025llm}. A recent study presented a promising approach to enhance the quality of next-step hint state-of-the-art methods by integrating LLMs with IDE internals such as static analysis and code quality checkers, improving the quality of next-step hints~\cite{birillo2024one}. The similar approach is widely used in other fields to enhance the output of models~\cite{blyth2025static, jaoua2025combining}. Since debugging is typically performed within IDEs, this method could potentially be combined with LLMs to assist students with debugging.

This paper introduces an AI-powered debugging assistant integrated into the open-source JetBrains Academy plugin for in-IDE learning~\cite{birillo2024bridging}. The tool provides real-time debugging support by analyzing students' code, suggesting breakpoints for buggy programs, and offering contextual explanations. It integrates Retrieval Augmented Generation (RAG)~\cite{gao2023retrieval} with LLMs to generate solutions and uses program slicing and custom heuristics to identify breakpoint locations efficiently. The source code of the tool can be found in the supplementary materials~\cite{supplementary}. A three-level evaluation, including technical analysis, UX evaluation, and classroom pilot tests, showed promising results, highlighting the tool's potential to assist students in debugging effectively.
\section{Related Work}
\label{section:related_work}

A common way to teach debugging is through debugging exercises~\cite{padurean2025bugspotter, whalley2021analysis, yang2025decoding}, where students identify and (sometimes) fix bugs in faulty code. 
Previous studies have expanded on this with approaches like 
scaffolding exercises with debug prints~\cite{fenwick2012using} or logs~\cite{chmiel2004debugging},
treating as a hypothesis-driven process~\cite{michaeli2019improving, whalley2021analysis} or adapting a troubleshooting framework~\cite{li2019towards}. 
Strategies also include learning materials, such as manuals~\cite{garcia2022read} or cheatsheets~\cite{ash2025wip}, as well as structured instruction, like debugging courses or interactive demonstrations~\cite{butler2025ildbug, wilkin2025debugging}. 
These approaches are effective, but often require extra materials and significant educator involvement for timely support.

The emergence of LLMs offers new opportunities to support students in debugging and practicing strategies. 
Chat-based assistants~\cite{kazemitabaar2024codeaid, liffiton2023codehelp} provide guarded feedback through conceptual explanations, pseudo code, and targeted annotations to help students find errors. 
Tools like \textit{BugSpotter}~\cite{padurean2025bugspotter} generate debugging tasks, while \textit{HypoCompass}~\cite{ma2024hypocompass} focuses on hypothesis construction through diagnosing buggy code. 
Despite their potential, LLM-based tools face challenges with reliability and seamless IDE integration.

To address this, some works embed debugging support into IDEs. 
For example, \textit{Ladebug}~\cite{luxton2018ladebug} provided a browser-based IDE-like environment where students practiced in predefined exercises. 
Recently, Noller et al. combined fault localization techniques with LLMs to guide students through the debugging process inside the IDE~\cite{noller2025simulated}, using automated breakpoints and a chatbot interface.  
While closely related, our work takes a different approach to breakpoint suggestion, focusing on performance, quality, and the cost of the solution. Additionally, we provide contextual hints for breakpoints directly within the editor, reducing reliance on a chatbot interface.
This distinction highlights that AI-powered debugging support can be integrated into IDEs in different ways, making it valuable to study a range of approaches.

Our work integrates debugging support into the IDE, aligning with students' workflows and reducing context switching. Unlike previous methods using predefined tasks, our tool works on students' code for personalized support. By combining IDE internals such as static analysis with LLMs, it provides reliable, scalable assistance without adding to educators' workload.

\section{AI-Debugging Tool}
\label{section:tool}

\subsection{Usage Pipeline}
\label{section:tool:usage_pipeline}

The AI-Debugging Assistant is integrated into the JetBrains Academy plugin for in-IDE learning~\cite{birillo2024bridging}. Students can view task descriptions and check their progress by clicking the \texttt{Check} button (\Cref{fig:tool:start_debug}-\textbf{1}), which runs tests locally. If an error happens (\Cref{fig:tool:start_debug}-\textbf{2}), the tool offers students the option to start a \textit{guided debugging session} (\Cref{fig:tool:start_debug}-\textbf{3}).

During the debugging session, the tool highlights recommended breakpoints in purple (\Cref{fig:tool:breakpoints}-\textbf{1}) to distinguish them from student-set breakpoints (\Cref{fig:tool:breakpoints}-\textbf{2}). Students follow these breakpoints and view explanation hints (\Cref{fig:tool:breakpoints}-\textbf{3}). After the session, the AI-recommended breakpoints convert to regular breakpoints, allowing students to continue the regular debugging process.

\begin{figure*}[t]
    \centering
    \includegraphics[width=\linewidth]{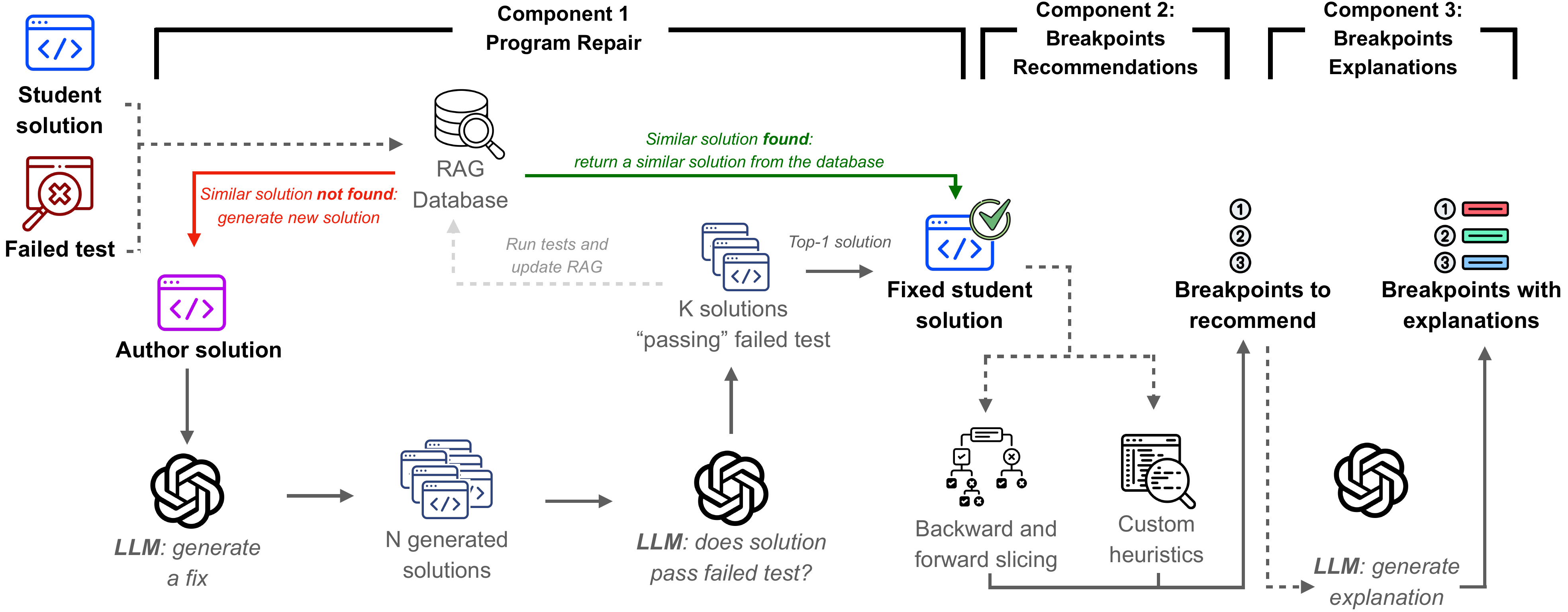}
    \caption{The tool pipeline: (1) program repair; (2) breakpoints recommendations; (3) breakpoints explanations components. The program repair component takes a student solution and a failed test, searching the RAG database for a similar correct solution. If a match is found, the process ends; if not, a new fixed solution is generated alongside the author’s initial solution. The breakpoint recommendations component analyzes the fixed solution, and the breakpoint explanations component generates explanations for each recommended breakpoint.}
    \label{fig:tool:pipeline}
\end{figure*}

\subsection{Tool Pipeline}
\label{section:tool:tool_pipeline}

The general idea of the approach is to generate a correct version of the student's solution that passes the failed test. Based on this, the AI-Debugging assistant generates and explains the breakpoints that should be shown to the student.
\Cref{fig:tool:pipeline} illustrates the tool's internal pipeline. The tool is composed of three main components: program repair (\Cref{fig:tool:pipeline}-\textbf{1}), breakpoints recommendations (\Cref{fig:tool:pipeline}-\textbf{2}), and breakpoints explanations (\Cref{fig:tool:pipeline}-\textbf{3}). Currently, the tool supports only Kotlin, but can be extended for other languages in the future.

\subsubsection{Program repair.} The first key component is program repair (\Cref{fig:tool:pipeline}-\textbf{1}), which takes the student's solution and the \textit{failed} test as input and returns a corrected solution that passes \textit{that} test. To generate a fix, we use a combined approach: retrieving a correct solution from existing ones using a RAG system~\cite{gao2023retrieval} or generating a new solution with an LLM. This minimizes unnecessary LLM calls and avoids hallucinations~\cite{zhang2025llm}, while remaining usable for new courses without prior student submissions.

\textbf{RAG system.} The RAG system uses existing solutions and information about tests, stored as embeddings generated by the \texttt{LaBSE} model\footnote{LaBSE model: \url{https://huggingface.co/sentence-transformers/LaBSE}}. The database can be empty (for new tasks) or pre-filled with previous submissions. When a debugging session is requested, embeddings are generated for the student’s solution and failed test. The system searches for a similar solution that passes the same test, using a cosine similarity threshold of $\geq 0.8$. If found, that solution is returned. Otherwise, the LLM generates a new solution.

\textbf{LLM solution generation.} The \texttt{\gptO} model is used to generate corrected solutions. Since LLMs are prone to hallucinations~\cite{smith2025spotting, koutcheme2025evaluating, chang2024survey, zhang2025llm}, five solutions are generated in parallel, and the best one is selected. Instead of running actual tests, the \texttt{o3-mini} model acts as a binary classifier to predict whether each solution will pass the failed test. Corrected solutions predicted to \textit{pass} are ranked using embeddings, prioritizing those closest to the student’s original code. The top-ranked solution is returned as the output.
The full prompts can be found in the supplementary materials~\cite{supplementary}.

\textbf{Uploading new solutions to the RAG system}. The LLM solution generation creates new submissions, which are added to the RAG system to expand the database and reduce future LLM calls. Initially, we rely on the model’s predictions to save time (\Cref{fig:tool:pipeline}-\textbf{2}). To ensure only valid solutions are added, we later execute actual tests.  This step not only enhances the database but also allows us to validate our approach and identify cases where the LLM's test failure predictions are inaccurate. This step is performed independently of the rest of the AI-Debugging assistant tool and does not affect its performance.

\subsubsection{Breakpoints recommendations.} When the correct version of the solution is generated, the next step is to suggest a \textit{set of possible breakpoints locations} to the student (\Cref{fig:tool:pipeline}-\textbf{2}). In our work, we consider two types of breakpoints: (1) breakpoints on the lines where the \textit{bug is located} and needs to be fixed directly, and (2) breakpoints on the lines that are \textit{likely affected by the bug}, which guide the student to the areas needing fixes. The first type of breakpoints is determined automatically by calculating the difference between the student's current solution and the correct one -- these are the exact locations that need to be changed for the solution to become correct. The more interesting part is using this information to find additional places the student should check to make these fixes. 

We use a combined approach to identify the lines that are likely affected by the bug. This includes programming \textit{backward and forward slicing}~\cite{korel1988dynamic, agrawal1990dynamic} as well as \textit{custom heuristics} to limit the number of recommended breakpoints. To show the final list of breakpoints to the student, we take the intersection of the breakpoints. We believe this approach works because it addresses two key problems: 1) it retains the most essential breakpoints, which are the most likely to impact the bug, and 2) it limits their number. This ensures that our system does not create an excessive number of breakpoints and allows the student to focus on debugging in the most relevant parts of the program.

\textbf{Backward and forward slicing}. Code slicing identifies code parts that influence (backward slice) or are influenced by (forward slice) a specific program element. It leverages static information to locate problem areas and is widely used in debugging tools~\cite{korel1988dynamic, agrawal1990dynamic}.

We built data and control flow graphs based on Program Structure Interface (PSI)\footnote{PSI: \url{https://plugins.jetbrains.com/docs/intellij/psi.html}} from the IntelliJ IDEA API to extract data control dependencies. \textit{Data dependencies} occur when the value of one variable depends on another, \textit{e.g.} in properties, or function calls. For example, in \texttt{var c = a + b}, \texttt{c} has backward dependencies on \texttt{a} and \texttt{b}, and \texttt{a} and \texttt{b} have forward dependencies on \texttt{c}. \textit{Control dependencies} occur when the execution of one statement depends on another, \textit{e.g.} in \texttt{if} statements or loops. For example, in an \texttt{if} statement, both conditional branches have backward control dependencies on the \texttt{if} statement.
As a result, this step produces forward and backward dependencies to the incorrect parts of the student's code. However, slicing alone may recommend too many lines, as it includes \textit{all} affected lines in the code. 

\textbf{Heuristics}. Based on our programming experience and previous research~\cite{fontana2021mapping}, we created a list of custom heuristics for possible places of breakpoints. We avoided using LLMs to ensure a deterministic approach, prevent model hallucinations, and reduce costs.

\begin{enumerate}
    \item \textbf{Conditional Statement Heuristic.} When the fix involves changes to a condition statement, \textit{e.g.}, \texttt{if}, \texttt{when}, or \texttt{while}, place breakpoints at the beginning of all code blocks that are executed based on these conditions. This allows monitoring how the logic flows under different scenarios.
    \item \textbf{Variable Modification Heuristic.} If the fix involves a line where a variable is modified, such as a property change, a binary expression, a loop parameter in a for expression, or a property in a while condition, place breakpoints to observe the state of the variable before and after its modification.
    \item \textbf{Function Scope Heuristic.} When the fix occurs within a function definition, place breakpoints at the beginning of the function to trace its execution and at all calls to the function in the code. This helps identify whether the function is invoked and behaves correctly after the change.
\end{enumerate}

\subsubsection{Breakpoints explanations.} The final step of the pipeline is the generation of breakpoints explanations (\Cref{fig:tool:pipeline}-\textbf{3}). We use \texttt{\gptO} to generate explanations for each breakpoint, helping students focus on specific code lines. The model identifies variables or objects being modified and provides detailed instructions for understanding these changes. For breakpoints on lines that are \textit{likely affected by the bug}, we request explanations about how those lines relate to the final error. For breakpoints on lines where the \textit{bug is located}, the model is instructed to explain why the error occurs~\cite{supplementary}.

\begin{table}[t]
    \centering
    \caption{Execution time comparison across candidate models, where P -- Precision, R -- Recall. Grey indicates the best model. }
    \label{tab:time-results}
    \begin{tabular}{lccccc}
    \toprule
    \textbf{Model} & \textbf{Avg (s)} & \textbf{Max (s)} & \textbf{P} & \textbf{R} & \textbf{F1} \\
    \midrule
    \texttt{\gptOMini (2048)} & 9.11 & 12.18 & - & - & -  \\
    \texttt{\gptOMini (1024)} & \textbf{8.24} & \textbf{10.68} & 0.68 & \textbf{0.88} & 0.76 \\
    \rowcolor{gray!20}\texttt{o3-mini (low)}              & \textbf{6.50} & \textbf{10.23} & \textbf{0.70} & 0.87 & \textbf{0.78} \\
    \texttt{o3-mini (medium)}           & 17.85 & 44.92 & - & - & - \\
    \bottomrule
\end{tabular}
\end{table}

\section{Evaluation}
\label{section:evaluation}

\subsection{RQ1: Breakpoints Recommendation Quality}
\label{section:evaluation:tech}

This section presents the technical evaluation of the proposed approach to answer \textit{RQ1: How precise is the AI-powered debugging system in recommending breakpoints?}. The evaluation focuses on the program repair (\Cref{fig:tool:pipeline}-\textbf{1}) and the breakpoints recommendation (\Cref{fig:tool:pipeline}-\textbf{2}) modules.

\subsubsection{Program Repair Evaluation}

The developed system relies on corrected student solutions that must pass the failed test. We use an LLM-based model to predict whether this solution will pass the failed test. To investigate how well such classifiers can replace actual test execution, we evaluated their feasibility as binary predictors of whether a generated fix passes the failed test.

\textbf{Execution time.}
We benchmarked four models on a sample of 150 solutions generated for the \textit{Kotlin Onboarding Introduction}\footnote{\label{onboarding}Kotlin Onboarding Introduction course: \url{https://plugins.jetbrains.com/plugin/21067-kotlin-onboarding-introduction}}. 
We ran the classifier on each solution, using a MacBook Pro (Apple M3 Max chip and 64 GB of memory). 
Based on the measurements (see \Cref{tab:time-results}), \texttt{o3-mini (low)} and \texttt{\gptOMini (1024)} models were selected for further evaluation because as the fastest ones.

\textbf{Quality evaluation.}
To evaluate how reliably the models can predict whether a generated fix is correct, we conducted an evaluation using $1,796$ incorrect Kotlin student submissions. 
For each submission, a fix was generated by \texttt{\gptO} (see \Cref{fig:tool:pipeline}-\textbf{1}). Then we ran two fastest models from the previous experiment on each fixed submission to predict if the fixed code is \textit{actually} fixes the student solution and validated the results by executing tests on each fixed submission. 
The performance of the classifiers is summarized in \Cref{tab:time-results}. \texttt{o3-mini} was chosen, balancing efficiency and quality. 

\subsubsection{Breakpoints Recommendation Evaluation}

Another key component of the developed system is the breakpoints recommendation module (\Cref{fig:tool:pipeline}-\textbf{2}). We assume the student's incorrect solution and the corresponding fixed solution are available. We evaluated the proposed approach, using a sample of 32 student solution pairs and their fixed versions from the Kotlin course\footref{onboarding}.
Two authors of this paper with over two years of research experience and more than four years of Kotlin programming experience independently labeled the data by identifying potential breakpoint locations. After labeling, they discussed the results to reach a consensus.
Using the labeled data, we evaluated the quality of the system. We achieved a Precision of $0.9$, Recall of $0.7$, and F1 score of $0.79$. These results demonstrate strong performance, so we decided to retain the current approach for further evaluation with students.

\subsection{RQ2: Tool's Usability}
\label{section:evaluation:ux}

The proposed system is embed to in-IDE environment, which might be hard for students to use~\cite{birillo2025ide}. To address this problem and answer \textit{RQ2: How do students perceive the tool's user experience, and how can it be improved?}, we conducted a usability study with students.

\textbf{Participants}. We invited eight CS students with basic Kotlin knowledge to participate in 15-minute semi-structured interviews. All participants had limited experience with debugging, \textit{i.e.}, they had tried using a debugger but did not use it regularly. 

\textbf{Experiment setting}. The experiment had two phases: (1) three pilot interviews, and (2) five interviews used for analysis. Interviews, conducted in English, used an interactive Figma prototype with limited functionality to allow interactions with the AI debugging assistant, but did not allow code editing. Participation was voluntary. We informed students about the study's purpose, data collection, and usage. 
Participants solved a beginner-level Kotlin problem with a buggy solution, using the prototype to locate the bug and explain how to fix it. The full interview script and the task can be found in the supplementary materials~\cite{supplementary}.

\textbf{Results}. Most participants found the tool intuitive after initial use. They appreciated the pre-configured breakpoints, which encouraged deeper exploration. Overall, the prototype was easy to use.
Many participants liked the idea of hints but found the content insufficient. Five noted that generic hints, like ``Check if this value is correct'' were insufficient. The final version was updated to include more detailed explanations.
Another issue noted by students was that breakpoint explanations overlapped with the code. In the initial prototype, hints were displayed by default. In the final version, hints now appear only when hovering over a breakpoint.

\subsection{RQ3: Student Perspective}
\label{section:evaluation:students}

The last step of evaluation was conducted as a pilot study in a classroom to answer \textit{RQ3: What are novice students' initial impressions of the tool in a classroom setting?}

\textbf{Participants.} We invited 20 1st- and 2nd-year Engineering Sciences bachelor's students. All participants had basic programming experience, having completed at least one semester of programming. To ensure clarity in the experiment, we selected students with limited debugging experience, assessed by their frequency of usage and self-reported confidence in the debugging process.

\textbf{Experiment setting.} The students participated in a 90-minute session conducted in a controlled lab environment. 
Participation was voluntary, and students were informed about the study’s purpose, data collection, and usage. No personal data was collected.
Students were provided with laptops preinstalled with the tool and printed instructions, including a description of the tool's concept and key components.
During the session, participants were tasked with two types of programming exercises -- several regular programming tasks requiring to implement some functions, and two debugging tasks -- one before and one after -- with multiple bugs in each~\cite{supplementary}. After completing the tasks, students filled out a survey.

\textbf{Results.} The experiments showed that only 8 out of 20 students used the AI Debugging Assistant during their debugging sessions. The most likely reason was a lack of Kotlin knowledge combined with the limited session time, which led to syntax errors and prevented students from fully utilizing the assistant. However, students were generally positive about the tool’s concept. 

According to results for the 1-5 Likert scale questions, the students who used the proposed system provided moderately positive feedback, with an average score of $3.13$ for the correctness of the provided breakpoint locations and $3.25$ for the clarity of the breakpoint explanations. In addition, three students highlighted that the tool's ability to localize and summarize the source of the bug was helpful for debugging. At the same time, two students mentioned that the automatically suggested breakpoints were useful and helped ease the debugging process.

Students suggested three main areas for improving the tool. First, four students recommended integrating the tool with a chat-based system to allow students to ask clarification questions. This idea could be easily implemented in future updates. Second, the user experience could be improved, as students found it unclear that they needed to hover over breakpoints to see explanation messages. This issue could likely be addressed with onboarding tools that demonstrate this feature during the first few uses. Lastly, some students mentioned adding a code generation feature to provide corrected solutions. However, this would require future investigation, as the tool's primary goal is to teach and assist with debugging rather than directly providing correct solutions.

\section{Threats to Validity}
\label{section:discussion}

The proposed approach has several key limitations. First, the technical evaluation was limited in scale. A greater diversity of tasks could enhance the custom heuristic set by identifying additional cases. This remains an area for future investigation.
Second, the pilot classroom evaluation involved a small number of participants. While 20 students were invited, only 8 used the tool during the session. Although we gathered initial evidence suggesting students appreciate the approach, we could not assess the actual learning impact of the tool, and results may also have been influenced by novelty effects. Future studies with a larger time for the experiment and more participants are planned to address this limitation. 
\section{Conclusion}
\label{section:conclusion}

This study addresses the gap in teaching debugging skills in CS curricula by presenting an AI-powered debugging assistant integrated into the IDE. The tool offers real-time, personalized debugging support by identifying breakpoints, providing explanations, and using techniques like RAG and program slicing to ensure accuracy and efficiency. Our three-level evaluation demonstrated the feasibility of this approach and suggested directions for future work on assessing its impact on student debugging practices and learning outcomes. This work provides a foundation for integrating real-time debugging tools into programming education and encourages further research on scalable solutions for teaching debugging.

\begin{acks}
We want to thank Evgenii Moiseenko for mentoring our team on this project, especially in programming slicing, code review, and discussing algorithms. A big thank you to Yaroslav Zharov and Rauf Kurbanov for their ML mentorship and guidance in building the system. We also thank the Human AI Experience team at JetBrains Research for designing and conducting the UX study with students, providing valuable insights to improve the system. Lastly, we thank the ETH Zurich Decision Science Laboratory for coordinating the student lab study.
\end{acks}

\bibliographystyle{ACM-Reference-Format}
\balance
\bibliography{ref}

\end{document}